\documentclass[]{spie}  %>>> use for US letter paper
\usepackage{amsmath,amsfonts,amssymb}
\usepackage{graphicx}
\usepackage{float}
\usepackage[utf8]{inputenc}
\usepackage[T1]{fontenc}
\usepackage[colorlinks=true, allcolors=blue]{hyperref}

\title{Automated brainstem parcellation using multi-atlas segmentation and deep neural network}

\author[a]{Magnus Magnusson}
\author[b,c]{Askell Love}
\author[*a,d]{Lotta M. Ellingsen}
\affil[a]{Dept. of Electrical and Computer Engineering, University of Iceland, Reykjavik, Iceland}
\affil[b]{Dept. of Medicine, University of Iceland, Reykjavik, Iceland}
\affil[c]{Dept. of Radiology, Landspitali – University Hospital, Reykjavik, Iceland}
\affil[d]{Dept. of Electrical and Computer Engineering, The Johns Hopkins University, Baltimore, MD, USA}

\authorinfo{*E-mail: lotta@hi.is}

\begin{document} 
\maketitle

\begin{abstract}
About 5-8$\%$ of individuals over the age of 60 have dementia. With our ever-aging population this number is likely to increase, making dementia one of the most important threats to public health in the 21st century. Given the phenotypic overlap of individual dementias the diagnosis of dementia is a major clinical challenge, even with current gold standard diagnostic approaches. However, it has been shown that certain dementias show specific structural characteristics in the brain. Progressive supranuclear palsy (PSP) and multiple system atrophy (MSA) are prototypical examples of this phenomenon, as they often present with characteristic brainstem atrophy. More detailed characterization of brain atrophy due to individual diseases is urgently required to select biomarkers and therapeutic targets that are meaningful to each disease. Here we present a joint multi-atlas-segmentation and deep-learning-based segmentation method for fast and robust parcellation of the brainstem into its four sub-structures, i.e., the midbrain, pons, medulla, and superior cerebellar peduncles (SCP), that in turn can provide detailed volumetric information on the brainstem sub-structures affected in PSP and MSA. The method may also benefit other neurodegenerative diseases, such as Parkinson’s disease; a condition which is often considered in the differential diagnosis of PSP and MSA. Comparison with state-of-the-art labeling techniques evaluated on ground truth manual segmentations demonstrate that our method is significantly faster than prior methods as well as showing improvement in labeling the brainstem indicating that this strategy may be a viable option to provide a better characterization of the brainstem atrophy seen in PSP and MSA.   
\end{abstract}


\section{INTRODUCTION}
\label{sec:intro}  % \label{} allows reference to this section

The brainstem has four main sub-structures, i.e., the midbrain, pons, medulla oblongata, and superior cerebellar peduncle (SCP). Recent research enabled by developments in MR imaging and image processing have begun to elucidate systematic relationships between area ratios and diameters of different brainstem sub-structures and the diagnoses of PSP and MSA\cite{Constantinides18}. These include methods to manually trace the midbrain and pons areas from the midsagittal slice followed by automated calculation of the midbrain to pons (M/P) area ratio\cite{silsby17}. Other methods measure the diameters of the middle cerebellar peduncle (MCP) and SCP to calculate the Magnetic Resonance Parkinsonism Index (MRPI)\cite{quattrone08}, MRPI=(P/M)x(MCP/SCP). Both the M/P ratio and the MRPI have shown to differ significantly in PSP compared with controls and the MRPI has been shown to correlate with the clinical evolution from Parkinson’s disease (PD) toward PSP\cite{silsby17,quattrone18}. Current neuroimaging correlates of MSA lack sufficient sensitivity and specificity to be used as reliable markers of MSA, however, current diagnostic criteria regard atrophy of pons as a supportive feature for possible MSA. These and other similar studies have two key limitations: First, most (although not all\cite{Iglesias15}) require manual delineations, limiting their use in large systematic research studies and clinical settings; and second, they only consider the area of the structures traced from a single midsagittal slice giving a limited and potentially inconsistent view of the three-dimensional (3D) brain structures.%sp\\

Recent advances in 3D image processing of MRIs have enabled the investigation of volumetric changes in the brainstem. One publication is available\cite{Iglesias15} on an algorithm performing automated segmentation and labeling of the brainstem into its four main sub-structures, which is now publicly available as a separate module in FreeSurfer\cite{Fischl12}. To get the brainstem sub-structures, one first needs to segment the whole brain using FreeSurfer's recon-all algorithm. This first step provides a segmentation of the whole brain, including the brainstem as one unit. Then the brainstem module is run to get the separate brainstem parcellation. Although highly successful in 3D segmentation, FreeSurfer’s key limitation is that it can take multiple hours to process a single scan, limiting its use in larger studies as well as in the clinical setting. Other brainstem parcellation approaches reported in the literature segment two or three sub-structures instead of the four main brainstem structures\cite{Nigro14,Sander19}. %\cite{Sander19}.%\\

Characteristic patterns of atrophy have been described in both PSP and MSA, however, for many of these patients, the current gold standard (i.e., radiologist’s evaluation) is rather nonspecific. Many overlapping clinical features are seen in PSP and MSA and this commonly leads to misdiagnosis with PD or Alzheimer’s disease. A major diagnostic challenge remains to identify and optimize the specific characteristics that can separate individual diseases at early stages, as only with robust early detection can investigators start testing possible therapeutic strategies and identify the underlying causes of these disorders. %\\

%\section{NEW WORK TO BE PRESENTED}
Here we propose a fast, accurate and robust 3D brainstem parcellation algorithm that addresses the key diagnostic imaging challenges by providing a detailed segmentation and labeling of the four brainstem sub-structures, i.e. the midbrain, pons, medulla oblongata, and SCP. The method uses a multi-atlas segmentation algorithm to automatically generate training data for a deep convolutional neural network (CNN), without the need for time consuming manual delineations of large training sets. The method provides significantly faster results than currently available state-of-the-art methods (less than one second rather than hours) for each new subject. This tool will enable faster and more accurate quantifications of brain atrophy in 3D that we hope will enable more robust differentiation of patients with PSP, MSA, and other neurodegenerative diseases at earlier stages than is currently possible.

\section{METHOD}

\subsection{Manual delineations}
\begin{figure} [ht]
   \begin{center}
   \begin{tabular}{c} %% tabular useful for creating an array of images 
   \centerline{\includegraphics[scale = 0.64]{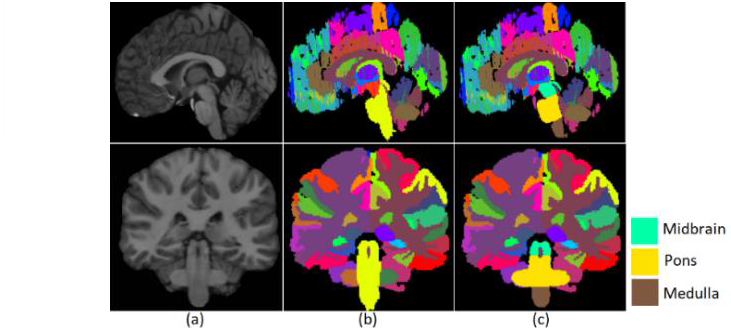}}
   \end{tabular}
   \end{center}
   \caption[example] 
%>>>> use \label inside caption to get Fig. number with \ref{}
   { \label{fig:delineation} 
The figure shows sagittal (top) and coronal (bottom) views of (a) T1-weighted MRI and (b) corresponding manual delineations from Neuromorphometrics; and (c) manual labeling of the midbrain, pons, and medulla according to our protocol.}
   \end{figure}

Detailed ground truth data is the key prior information for automated multi-atlas segmentation methods as well as being critical for the evaluation of automated labeling methods. For developmental purposes we used imaging data from Neuromorphometrics (www.neuromorphometrics.com), comprising T1-weighted MRIs and corresponding atlas images, each with 134 manually labeled anatomical structures [see Figure \ref{fig:delineation}(a)-(b)]. In these images the brainstem was labeled as a single unit. We developed a new brainstem delineation protocol for detailed and robust parcellation of the brainstem into midbrain, pons, medulla, and SCP, for the two main purposes of 1) having a strict guideline to ensure that each scan would be labeled in the same fashion to maximize robustness of the automated segmentation method; and 2) to have detailed and easy to follow guidelines for future manual delineations of new brain scans. The delineation protocol was based on the protocol in \cite{Iglesias15} and expert guidance from an experienced neuroradiologist. A total of 15 MRIs were manually labeled according to our new delineation protocol [see Figure \ref{fig:delineation}(c)] to be used as atlases in our multi-atlas segmentation approach. We used the Medical Image Processing, Analysis, and Visualization tool (MIPAV, see mipav.cit.nih.gov) for manual labeling.

\subsection{Multi-atlas segmentation and generation of training data}
\label{sec:Multi}

One of the main requirements for the development of a robust and accurate deep learning-based segmentation method is having a rich set of labeled training data. Manual segmentation is still the current gold standard for brain images; however, they are laborious and time consuming to generate. Here we utilized the multi-atlas-segmentation framework of the robust dictionary learning and label propagation hybrid (RUDOLPH) method\cite{Ellingsen16} to automatically segment the four brainstem sub-structures. We incorporated the 15 subjects from Neuromorphometrics that were manually labeled according to our delineation protocol into the RUDOLPH framework and adjusted the algorithm to account for the new brainstem labels. This provided us with automated labeling of the sub-structures of the brainstem to be used as training data for our CNN. %\\

% Note: If compiling with LaTeX+dvipdf, please ensure images generated from 
% other software packages have their bounding boxes set correctly.

\newpage
We applied the augmented RUDOLPH method to 35 MRIs from Neuromorphometrics to automatically generate training data for our deep neural network. The MRIs were cropped to the size of $96\times96\times96$ voxels, from the original size of $240\times285\times240$ voxels, to ensure that each image would only contain the brainstem and nearby structures and to speed up the execution of the neural network. Finally, the brainstem labels were extracted from these results to produce the final training set comprising five labels, i.e. the midbrain, pons, medulla, SCP, and background. Using this approach enabled us to automatically generate our training set and avoid the time-consuming work of manually labeling multiple training images.

\subsection{Network architecture and training procedure}

\begin{figure} [ht]
   \begin{center}
   \begin{tabular}{c} %% tabular useful for creating an array of images 
   \includegraphics[scale = 0.45]{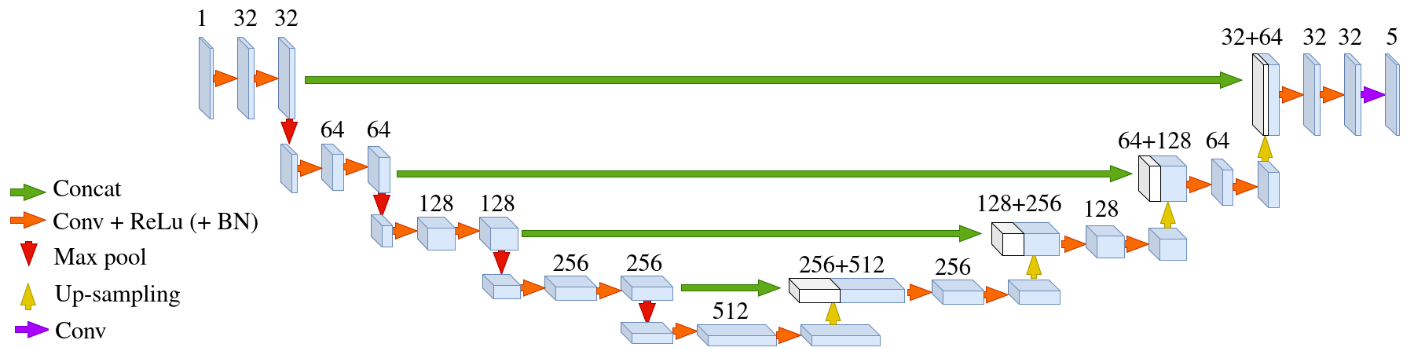}
   \end{tabular}
   \end{center}
   \caption[example] 
%>>>> use \label inside caption to get Fig. number with \ref{}
   { \label{fig:tauganet} 
A schematic of the 3D U-net architecture of the proposed method.}
   \end{figure}

 Our model was based on the 3D U-Net architecture \cite{Ronneberger15,CicekALBR16}. A schematic of the network is shown in Figure \ref{fig:tauganet}. We used five resolution steps, where each step in the contracting path consisted of two $3\times3\times3$ convolutional layers, with rectified linear unit (ReLu) activation\cite{NairH10}, and a batch normalization\cite{pmlr-v37-ioffe15} layer after the latter convolutional layer. The output of each batch normalization layer was fed to the corresponding resolution step in the expanding path, and to a $2\times2\times2$ max pooling layer in the contracting path. The last layer in the contracting path only consisted of two convolutional layers, without batch normalization or max pooling; its output was then upsampled and fed to the expanding path. The expanding paths steps mirrored the contracting path. Its inputs came from the shortcut connection, after each batch normalization layer, which were concatenated with $2\times2\times2$ upsampled inputs from the previous step. This was followed by two $3\times3\times3$ convolutional layers with ReLu activation. The final output was a $1\times1\times1$ convolutional layer, with a 5 class softmax activation. 
 %\newpage
   \begin{figure} [!ht]
   \begin{center}
   \begin{tabular}{c} %% tabular useful for creating an array of images 
   \centerline{\includegraphics[scale = 0.5]{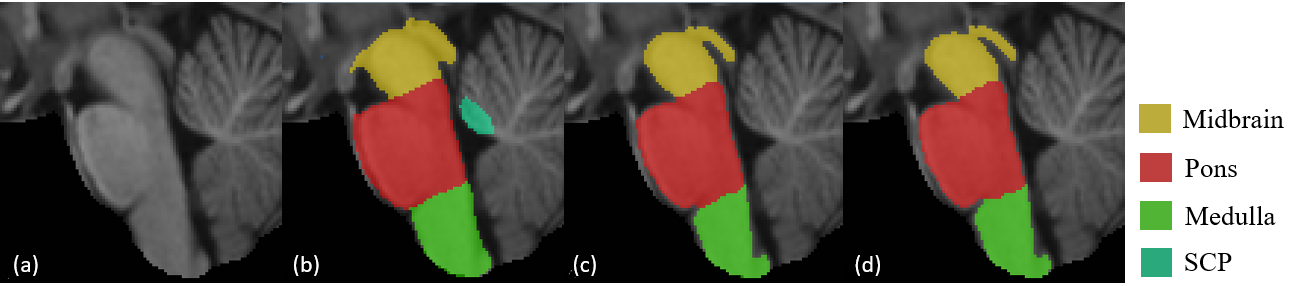}}
   \end{tabular}
   \end{center}
   \caption[example] 
%>>>> use \label inside caption to get Fig. number with \ref{}
   { \label{fig:loss} 
Brainstem segmentation results of a single subject using our model with different loss functions: (a) Dice loss only; (b) weighted cross-entropy loss only; (c) Dice Loss using pre-trained weights from (b). (d) The manually labeled mask shown for comparison. Only using the Dice loss resulted in dead ReLUs so no training occurred, while the weighted cross-entropy loss oversegmented the MRIs.}
   \end{figure}
   
 The intensity values of the MRIs were normalized to [0, 1]. The set of 35 RUDOLPH processed MRIs were split into 27 training and 8 validation images. The original 15 manually delineated MRIs were used as test images. Each voxel was one-hot-encoded, corresponding to one of five categories: 1. Background $\simeq
 91.6\%$; 2. Pons $\simeq 5\%$; 3. Midbrain $\simeq 2.25\%$; 4. Medulla $\simeq 1.1\%$ ; and 5. SCP $\simeq 0.08\%$ of the data set. The brainstem labeled voxels covered less than $10 \% $ of the entire data set. Class imbalance is a common problem in biomedical image segmentation, where small structures need to be segmented from a much larger background. This problem was addressed by training our model using two loss functions. First, we pre-trained the model using a softmax weighted cross-entropy loss\cite{Ronneberger15}. Second, we initialized our final model using the pre-trained weights and trained using the Dice loss function \cite{milletari2016v}, since neither loss function gave adequate results when used by itself [see Figure~\ref{fig:loss}(a)-(b)]. The model achieved high accuracy scores and low loss when combining pre-trained weights and the Dice loss [see Figure \ref{fig:loss}(c)]. %\\
 
 The model was pre-trained for 20 epochs and trained for 200 epochs using the final model. A batch size of 1 was used both in pre-training and training. Stochastic gradient descent with a learning rate of 0.01 and momentum of 0.9 was used for optimization. No data augmentation was used.

\section{RESULTS}

\begin{figure} [H]
   \begin{center}
   \begin{tabular}{c} %% tabular useful for creating an array of images 
   \includegraphics[scale = 0.646]{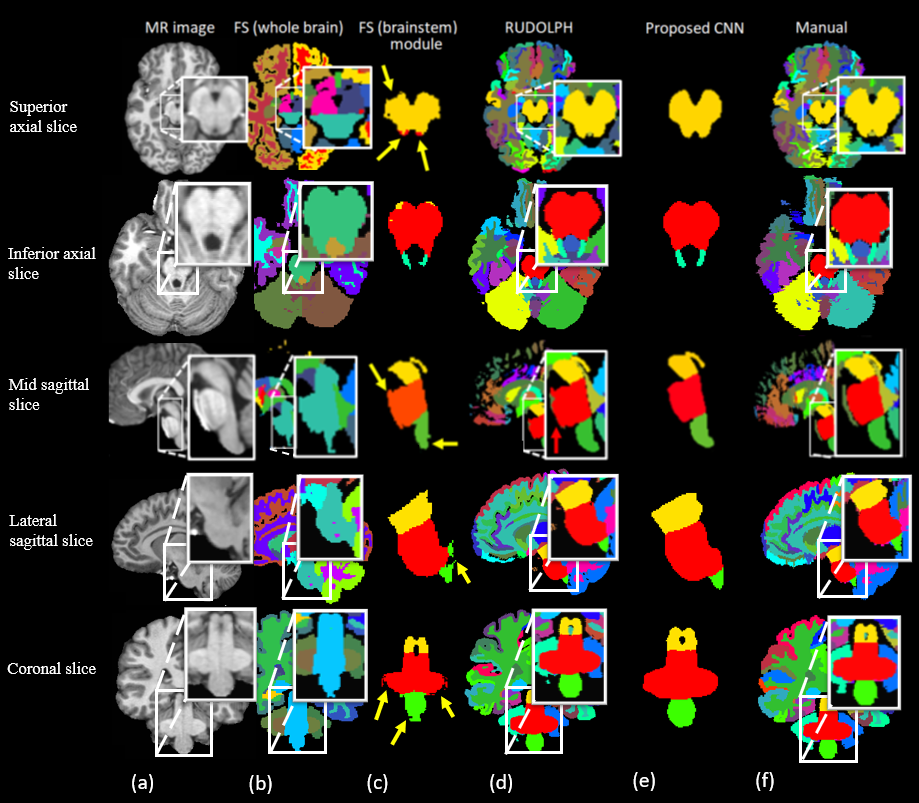}
   \end{tabular}
   \end{center}
   \caption[example] 
%>>>> use \label inside caption to get Fig. number with \ref{}
   { \label{fig:FSvsRvsCNN} 
Brainstem segmentation results of five subjects from (b) FreeSurfer (FS) (whole brain); (c) FreeSurfer -brainstem module (see yellow arrows for errors); (d) RUDOLPH using a 14-to-1 crossvalidation scheme (see red arrow for error); (e) our proposed CNN; compared with (f) manual rater for (a) the subject's T1-weighted MRI. The white boxes show the zoomed in brainstem regions. Color code for the brainstem sub-structures in (c-f): Midbrain (yellow), pons~(red), medulla (green), SCP (cyan).}
   \end{figure}

We evaluated the proposed method on 15 manually segmented images (the current gold standard) using the Dice similarity coefficient (DSC)\cite{dice} and compared our method with the state-of-the-art segmentation method FreeSurfer in addition to our augmented RUDOLPH method. A visual comparison is shown in Figure \ref{fig:FSvsRvsCNN}, indicating improved segmentation of our proposed CNN over FreeSurfer and RUDOLPH. Visually, our proposed CNN segments the sub-structures of the brainstem accurately, when compared to our manually delineated mask. However, we note that there were inconsistencies between the segmentation of the whole brain module and the brainstem module of FreeSurfer. Large parts of the midbrain and the pons were not labelled as part of the brainstem in the whole brain module [see Figure \ref{fig:FSvsRvsCNN} (b) superior axial, sagittal, and coronal slices]. The brainstem module also produced pixelated and disconnected segmentations shown with yellow arrows in the lateral sagittal and coronal slices [see Figure \ref{fig:FSvsRvsCNN} (c)].

The accuracy of the segmentation was evaluated quantitatively by calculating the DSC, between the predicted segmentation from the CNN and the original manually delineated masks. The DSC measures the overlap between two binary mask, X and Y, and is calculated using Eq.  \ref{eq:dsc}.

\begin{align}
    \mathrm{DSC}=\frac{2|X\cap Y|}{|X| + |Y|} \label{eq:dsc} %\frac{2TP}{2TP + FP + FN}
\end{align}
\newpage

\noindent The range of DSC is [0,1], where 1 means perfect overlap and 0 means no overlap between the two masks, X and Y. The average DSC for the brainstem CNN evaluated on 15 manually labeled subjects was: Pons: 0.96$\pm$0.0059; midbrain: 0.93$\pm$0.015; medulla: 0.93$\pm$0.009; and SCP: 0.79$\pm$0.039. The DSC is high for the pons, midbrain, and medulla oblongata. This indicates that the model accurately segments these structures. The DSC is lower for the SCP, compared to the other structures, which can be explained by the size of the structure, as smaller structures tend to have a lower DSC. We did not compute the DSC for FreeSurfer based on our manual masks, since the two methods are based on slightly different delineation protocols. However, we note that our DSC is substantially higher than previously reported numbers for FreeSurfer \cite{Iglesias15}. Our method also yields a significant speedup, segmenting each new subject in 0.5s compared to 8-12h using RUDOLPH or FreeSurfer.%\\

\section{Discussion and Conclusion}
We present a new method for automated parcellation of the brainstem into its four sub-structures, i.e. midbrain, pons, medulla oblongata, and SCP, using a 3D U-net model. The method was trained on 35 labeled training images automatically generated by a multi-atlas segmentation method, thereby eliminating the need for manually delineated training data. The method was compared with state-of-the-art segmentation methods and evaluated on ground truth manual labels. The average DSC over 15 subjects was: Pons: 0.96$\pm$0.0059; midbrain: 0.93$\pm$0.015; medulla: 0.93$\pm$0.009; and SCP: 0.79$\pm$0.039. The method also provides significant speedup compared with current methods, segmenting each new scan in only 0.5s. Future work includes further evaluation of the method on a larger cohort of healthy subjects and patients with PSP and MSA. It is our hope that the proposed method will provide researchers with the means to quantify the brainstem atrophy seen in PSP and MSA with higher specificity than current approaches can provide. 

\acknowledgments
This work was supported by the RANNIS Icelandic Student Innovation Fund.
% References
\bibliography{main} % bibliography data in main.bib
\bibliographystyle{spiebib} % makes bibtex use spiebib.bst

\end{document}